\documentclass[pre,aps,twocolumn, superscriptaddress]{revtex4}
\usepackage{amsmath}
\usepackage{epsfig}
\usepackage{amssymb}
\usepackage{color}
\usepackage{caption}
\begin{document}

\title{Langevin thermostat for robust configurational and kinetic sampling}

\author{Oded Farago}
\affiliation{Department of Chemistry, University of Cambridge,
  Lensfield Road, Cambridge CB2 1EW, United Kingdom}
\affiliation{Department of Biomedical Engineering, Ben-Gurion
  University of the Negev, Be'er Sheva 85105, Israel}

\begin{abstract}

We reformulate the algorithm of Gr{\o}nbech-Jensen and Farago (GJF)
for Langevin dynamics simulations at constant temperature. The GJF
algorithm has become increasingly popular in molecular dynamics
simulations because it provides robust (i.e., insensitive to
variations in the time step) and accurate configurational
sampling of the phase space with larger time steps than other Langevin
thermostats. In the original derivation [Mol. Phys. {\bf 111}, 983
  (2013)], the algorithm was formulated as a velocity-Verlet type
integrator with an in-site velocity variable. Here, we
reformulate it as a leap frog scheme with a half-step velocity
variable. In contrast to the original form, the
reforumlated one also provides robust and accurate estimations of
kinetic measures such as the average kinetic energy. We analytically
prove that the newly presented algorithm gives the exact
configurational and kinetic temperatures of a harmonic oscillator for
any time step smaller than the Verlet stability limit, and use
computer simulations to demonstrate the configurational and kinetic
robustness of the algorithm in strongly non-linear systems. This
property of the new formulation of the GJF thermostat makes it very
attractive for implementation in computer simulations.

\end{abstract} 

\maketitle

\section{Introduction}

One of the prominent approaches for conducting molecular simulations
in the canonical $(N,V,T)$ ensemble is based on the idea that the
statistical ensemble can be sampled by considering the dynamics of
each particle in the system to be governed by Langevin
equation~\cite{langevin:1908}
\begin{equation}
  m\ddot{r}=f-\alpha v+\beta(t),
  \label{eq:langevin}
\end{equation}
where $r$ and $v=\dot{r}$ denote, respectively, the coordinate and
velocity of the particle. Langevin's equation is essentially Newton's
second law describing the motion of a particle of mass $m$ under the
action of (i) a deterministic force, $f$, and two additional forces
representing the interactions with a heat bath - (ii) a friction
force, $-\alpha v$, where $\alpha>0$ is the friction coefficient, and
(iii) Gaussian white noise with zero mean and delta-function
auto-correlation~\cite{risken:book}:
\begin{eqnarray}
  \langle\beta(t)\rangle&=&0\\
  \langle\beta(t)\beta(t^{\prime})\rangle&=&2k_BT\alpha\delta(t-t^{\prime}),
\end{eqnarray}
where $k_B$ is Boltzmann's constant and $T$ is the temperature of the
heat bath.

In computer simulations the time is discretized in intervals of $dt$,
and a Langevin ``thermostat'' algorithm is used for discrete-time
integration of Langevin's equation of motion, yielding a sequence of
coordinates $r^n=r(t_n)$ and velocities $v^n=v(t_n)$, where
$t_n=ndt$. A major problem of thermostat algorithms is the
discretization errors that they introduce~\cite{pbs88}. These cause
computed averages of thermodynamic quantities of interest to vary with
the time step $dt$ - an alarming feature that raises concerns about
the reliability of the simulation results. A simple test-case for the
robustness of an algorithm is the one-dimensional harmonic oscillator
where $f=-kr$. The average potential energy satisfies $\langle
E_p\rangle=\langle kr^2/2\rangle=k_BT/2$, but most popular and
widely-implemented algorithms, e.g., BBK~\cite{bbk84},
Schneider-Stoll~\cite{ss78}, and van Gunsteren-Berendsen~\cite{vGB82}
thermostats, exhibit systematic deviations from this result. Depending
on the method of choice, the integration error of the potential energy
may scale as ${\cal O}(dt)$ or ${\cal O}(dt^2)$~\cite{wang03}. It was
only several years ago that two new algorithms were introduced, by
Leimkuhler and Matthews (LM)~\cite{lm12} and by Gr{\o}nbech-Jensen and
Farago (GJF)~\cite{gjf1}, that reproduce the exact harmonic oscillator
potential energy for any time step within the stability limit
$dt<2/\Omega_0$, where $\Omega_0=\sqrt{k/m}$ is the frequency of the
oscillator.

When conducting a molecular simulations study, one is often interested
in measuring the temperature of the simulated system in order to
compare it to the target thermodynamic temperature. The most
straightforward quantity to calculate for this purpose is the average
kinetic energy per degree of freedom, $\langle E_k\rangle=\langle
mv^2/2\rangle=k_BT/2$.  Unfortunately, the discrete-time variables
$r^n$ and $v^n$ are only approximations of their continuous-time
counterparts. In contrast to the latters, the formers are not exactly
conjugated to each other, which causes the ``kinetic'' and
``configurational'' measures of the temperature to be different. This
feature is nicely captured by the harmonic oscillator test-case. As
mentioned above, the LM and GJF algorithms yield the correct
configurational temperature,
\begin{equation}
  \left\langle E_p\right\rangle
  =\left\langle\frac{k\left(r^n\right)^2}{2}\right\rangle=\frac{k_BT}{2},
    \label{eq:epotential}
\end{equation}
but the kinetic temperature computed by these thermostats exhibits a
discretization error and reads
 \begin{equation}
   \left\langle E_k\right\rangle
   =\left\langle\frac{m\left(v^n\right)^2}{2}\right\rangle=
  \frac{k_BT}{2}\left[1-\frac{\left(\Omega_0 dt\right)^2}{4}\right].
  \label{eq:ekinetic}
 \end{equation}
There exist other thermostats that reproduce the kinetic energy
without discretization errors, but no existing algorithm has
simultaneously {\em both}\/ the correct kinetic and potential energy
of the harmonic oscillator. Since the aim of computer simulation
studies of molecular systems at equilibrium is phase space sampling,
the velocity variable is essentially an auxiliary field and one should
favor the use of algorithms like the GJF thermostat, which have been
demonstrated to provide robust configurational sampling not only for
the harmonic oscillator but also for non-linear molecular
systems~\cite{gjf2,gjf3}. Nevertheless, the kinetic energy constitutes
a useful and a simple measure for the temperature of the system and,
therefore, a question arises on whether it is possible to devise a
thermostat featuring both correct potential and kinetic energies of
the harmonic oscillator. Here, we show that the GJF algorithm can be
reformulated with a different velocity variable which, in contrast to
the one in the original formulation, exhibits no discretization
errors. We use simulations of a simple toy model to demonstrate the
robustness of the newly-defined velocity also in non-linear systems.

\section{Half-step velocity}
\label{sec:halfstep}

Our starting point is the GJF algorithm, which in the velocity-Verlet
formulation reads~\cite{gjf1}
\begin{eqnarray}
  r^{n+1}&=&r^n+b\left[v^ndt+\frac{dt^2}{2m}f^n+\frac{dt}{2m}\beta^{n+1}\right]
  \label{eq:gjfr}\\
  v^{n+1}&=&av^n+\frac{dt}{2m}\left(af^n+f^{n+1}\right)+\frac{b}{m}\beta^{n+1},
  \label{eq:gjfv}
\end{eqnarray}
where $f^n=f(r^n)$, and the damping coefficients of the algorithm are given by
\begin{equation}
  a=\frac{1-\frac{\alpha dt}{2m}}{1+\frac{\alpha dt}{2m}}
  \label{eq:acoefficient}
\end{equation}
and 
\begin{equation}
  b=\frac{1}{1+\frac{\alpha dt}{2m}}.
  \label{eq:bcoefficient}
\end{equation}
The discrete-time noise,
\begin{equation}
  \beta^{n+1}=\int_{t_n}^{t_{n+1}}\beta(t^{\prime})dt^{\prime},
\end{equation}
is a random Gaussian number satisfying
\begin{eqnarray}
  \left\langle \beta^n\right\rangle&=&0\\
  \left\langle \beta^m\beta^n\right\rangle&=&2k_BT\alpha dt\delta_{m,n}
  \label{eq:beta2}
\end{eqnarray}
where $\delta_{m,n}$ is Kronecker delta.

We now invoke another property of the canonical ensemble, which is the
fact that $r$ and $v$ are statistically independent degrees of
freedom, namely $\langle rv\rangle=0$. Let us demonstrate that the
discrete-time variables in the GJF algorithm satisfy this relation. To
show this, we begin by squaring Eq.~(\ref{eq:gjfr}) and taking
statistical averages of all terms. Keeping also in mind that
$f^n=-kr^n$, this yields
\begin{widetext}
\begin{eqnarray}
 \left\langle\left(r^{n+1}\right)^2\right\rangle=
 \left\langle\left(r^n\right)^2\right\rangle
 \left[1-\frac{b\left(\Omega_0dt\right)^2}{4}\right]^2
 +b^2dt^2\left\langle\left(v^n\right)^2\right\rangle
 +\frac{b^2dt^2}{4m^2}\left\langle\left(\beta^{n+1}\right)^2\right\rangle
 +2\left\langle
 r^nv^n\right\rangle\left[1-\frac{b\left(\Omega_0dt\right)^2}{4}\right]
 bdt.
 \label{eq:rv}
\end{eqnarray}
\end{widetext}
Using Eqs.~(\ref{eq:epotential}), (\ref{eq:ekinetic}), and
(\ref{eq:beta2}), and the fact that
$\left\langle\left(r^{n+1}\right)^2\right\rangle
=\left\langle\left(r^n\right)^2\right\rangle$
in Eq.~(\ref{eq:rv}), yields the equality
\begin{equation}
  \left(\Omega dt\right)^2\left(b^2-b+\frac{\alpha dt}{2m}\right)+
  2\left\langle
  r^nv^n\right\rangle\left[1-\frac{b\left(\Omega_0dt\right)^2}{2}\right]dt=0.
  \label{eq:equality1}
\end{equation}
The first term on the r.h.s.~of Eq.~(\ref{eq:equality1}) vanishes
because $b^2-b+\frac{\alpha dt}{2m}=0$, and we immediately conclude
that
\begin{equation}
  \left\langle r^nv^n\right\rangle=0.
  \label{eq:rv2}
\end{equation}

Let us look at the ``half-step'' velocity variable, $u^{n+1/2}$,
defined by rewriting the {\em frictionless\/} ($\alpha=0$)
velocity-Verlet algorithm~\cite{verlet67} in the following form:
\begin{eqnarray}
  u^{n+\frac{1}{2}}&=&v^{n}+\frac{dt}{2m}f^{n}
  \label{eq:vvv2half}\\
  r^{n+1}&=&r^n+u^{n+\frac{1}{2}}dt
  \label{eq:vvr2}\\
  v^{n+1}&=&u^{n+\frac{1}{2}}+\frac{dt}{2m}f^{n+1}.
  \label{eq:vvv2}
\end{eqnarray}
With the definition of $u^{n+1/2}$ by Eq.~(\ref{eq:vvv2half}), the GJF
equations (\ref{eq:gjfr})-(\ref{eq:gjfv}) in the velocity-Verlet form,
can be converted into the following leap-frog form
\begin{eqnarray}
    u^{n+\frac{1}{2}}&=&au^{n-\frac{1}{2}}+\frac{dt}{m}f^{n}
  +\frac{b}{m}\beta^{n}
  \label{eq:gjfv2}\\
  r^{n+1}&=&r^n+b\left[u^{n+\frac{1}{2}}dt+\frac{dt}{2m}\beta^{n+1}\right].
  \label{eq:gjfr2}
\end{eqnarray}
These equations constitute a new formulation of the GJF thermostat, to
be henceforth referred to as the GJF-F algorithm.  The half-step
velocity variable $u^{n+1/2}$ satisfies
\begin{eqnarray}
  \left\langle\left(u^{n+\frac{1}{2}}\right)^2\right\rangle&=&
  \left\langle\left(v^n+\frac{f^n}{2m}\right)^2\right\rangle
  \label{eq:u2}\\
  =\left\langle\left(v^n\right)^2\right\rangle
  &+&\frac{k^2dt^2}{4m^2}\left\langle\left(r^n\right)^2\right\rangle
  -\frac{kdt}{m}\left\langle r^nv^n\right\rangle,\nonumber
\end{eqnarray}
and using Eqs.~(\ref{eq:epotential}), (\ref{eq:ekinetic}), and
(\ref{eq:rv2}) we readily find the kinetic energy associated with
$u^{n+1/2}$
 \begin{equation}
   \left\langle E_k\right\rangle=
   \left\langle\frac{m\left(u^{n+\frac{1}{2}}\right)^2}{2}
   \right\rangle=\frac{k_BT}{2},
  \label{eq:ekinetic2}
\end{equation} 
which is exact for any time step $dt$ (within the stability limit).

With the above derivation, it is easy to define another half-step
velocity
\begin{equation}
  w^{n+\frac{1}{2}}=\frac{r^{n+1}-r^n}{\sqrt{b}dt}
  \label{eq:uhalf}
\end{equation}
with similar properties. This velocity variable was independently
identified recently by Gr{\o}nbech-Jensen and Gr{\o}nbech-Jensen
(2GJ)~\cite{2gj}. In order to prove that $w^{n+1/2}$ is a robust
velocity variable, we rewrite Eq.~(\ref{eq:gjfr2}) in a slightly
different form
\begin{equation}
  r^{n+1}-r^n=b\left[u^{n+\frac{1}{2}}dt+\frac{dt}{2m}\beta^{n+1}\right].
  \label{eq:gjfr3}
\end{equation}
Squaring both sides of Eq.~(\ref{eq:gjfr3}) and taking averages, we
arrive at
\begin{equation}
  \left(r^{n+1}-r^n\right)^2=b^2dt^2\left[\left\langle
    \left(u^{n+\frac{1}{2}}\right)^2\right\rangle+\frac{1}{4m^2}
    \left\langle\left(\beta^{n+1}\right)^2\right\rangle\right],
  \label{eq:equality2}
\end{equation}
and by using Eqs.~(\ref{eq:ekinetic2}) and (\ref{eq:beta2}) we find that
\begin{equation}
  \left(r^{n+1}-r^2\right)^2=b^2dt^2\frac{k_BT}{m}\left[1+\frac{\alpha
      dt}{2m}\right]=bdt^2\frac{k_BT}{m}.
  \label{eq:equality3}
\end{equation}
From the last result we immediately conclude that for any $dt$
\begin{equation}
   \left\langle E_k\right\rangle=
   \left\langle\frac{m\left(w^{n+\frac{1}{2}}\right)^2}{2}\right\rangle=
     \frac{k_BT}{2}.
  \label{eq:ekinetic3}
\end{equation}
A leap-frog scheme involving $w^{n+1/2}$ can be derived by
complementing Eq.~(\ref{eq:uhalf}) with the Str{\o}mer-Verlet form of
the GJF algorithm (see Eq.~(11) in ref.~\cite{gjf2})
\begin{equation}
  r^{n+1}=2br^n-ar^{n-1}+\frac{bdt^2}{m}f^n+\frac{bdt}{2m}
  \left(\beta^n+\beta^{n+1}\right),
  \label{eq:svgjf}
\end{equation}
which, together with the relationship $a+1=2b$, leads to the following
set of equations
\begin{eqnarray}
  w^{n+\frac{1}{2}}&=&aw^{n-\frac{1}{2}}+\frac{\sqrt{b}dt}{m}f^n+
  \frac{\sqrt{b}}{2m}\left(\beta^n+\beta^{n+1}\right)
  \label{eq:gjfvlf}\\
  r^{n+1}&=&r^n+\sqrt{b}w^{n+\frac{1}{2}}dt\label{eq:gjfrlf}.
\end{eqnarray}
This scheme was presented in ref.~\cite{2gj} and was termed the
GJF-2GJ algorithm.

\section{Simulations of a non-linear model}

\begin{figure*}
\includegraphics[width=1\textwidth]{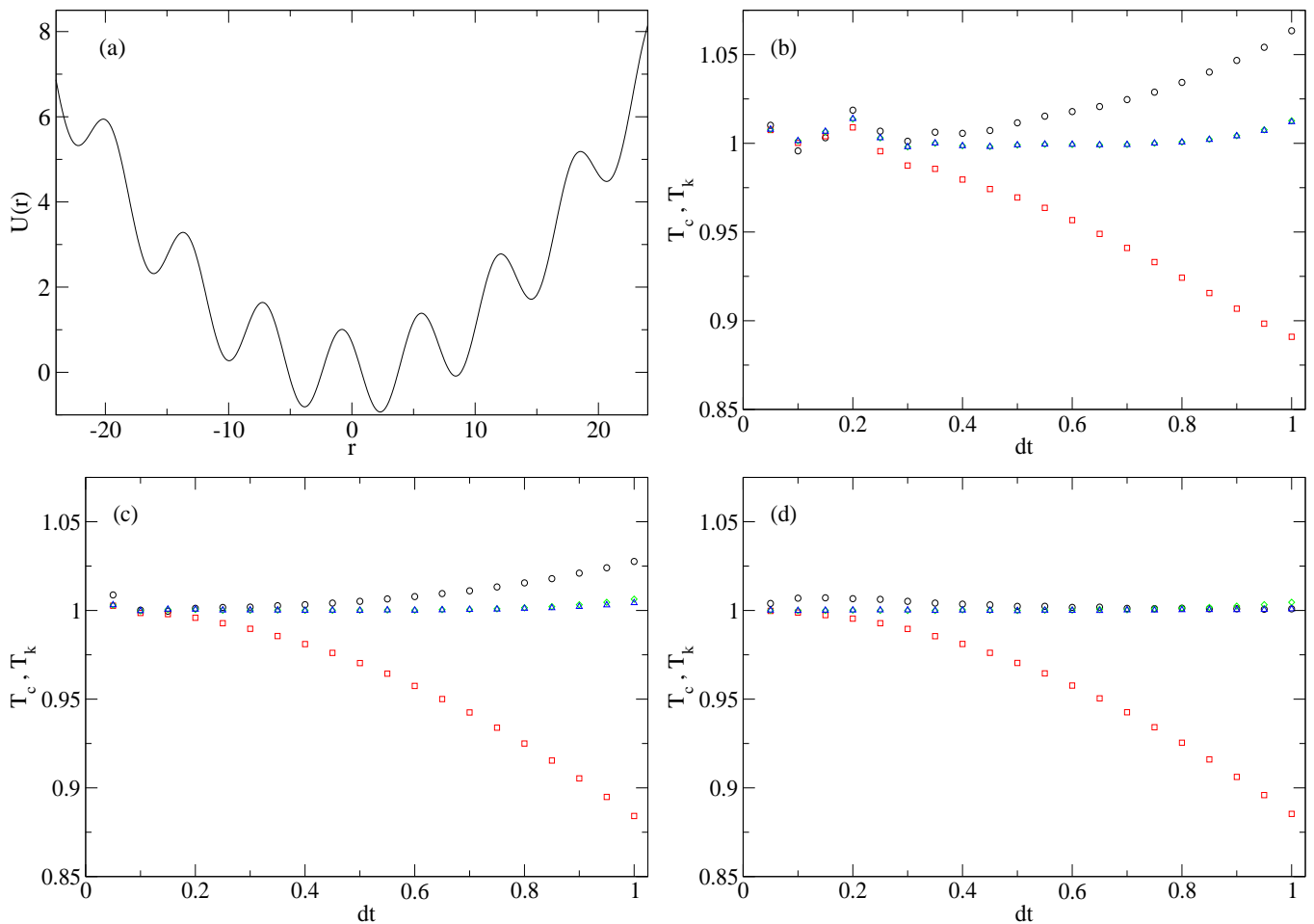}
\caption{(a) The simulated anharmonic potential energy function
  $U(r)=kr^2/2-\cos(r-\xi)$, with $k=1/40$, and $\xi=3/4\pi$. (b-d)
  The configurational temperature, $T_c$ (black circles), and the
  kinetic temperatures, $T_k^v$ (red squares), $T_k^u$ (green
  diamonds), and $T_k^w$ (blue triangles), as a function of the
  simulation time step $dt$. The simulation friction coefficient is
  $\alpha=0.1$ in (b), $\alpha=1$ in (c), and $\alpha=10$ in (d).}
\label{fig1}
\end{figure*}

To test the robustness of the new velocity variables beyond the
harmonic oscillator test case, we consider the non-linear model
presented in ref.~\cite{gjf3} of a particle moving in a
one-dimensional potential $U(r)=kr^2/2-\cos(r-\xi)$, with $k=1/40$,
and $\xi=3/4\pi$ [see fig.~\ref{fig1}(a)]. In ref.~\cite{gjf3}, this
model provided a demonstration for the superiority of the GJF
algorithm over classical popular thermostats (BBK, SS, vGB) in
configurational sampling. This was done by computing the
configurational temperature defined by
\begin{equation}
  T_c=\frac{1}{k_B}\frac{\left\langle\left(\frac{\partial U}{\partial r}
    \right)^2\right\rangle}
  {\left\langle\frac{\partial^2 U}{\partial r^2}\right\rangle}.
    \label{eq:tc}
\end{equation}
We now wish to also measure the kinetic temperature
\begin{equation}
  T_k=\frac{2}{k_B}\left\langle E_k\right\rangle,
  \label{eq:tk}
\end{equation}
and explore the dependence of this quantity of the simulation time
step $dt$. Our simulation results for the dependence of the
configurational and kinetic temperatures on $dt$ are summarized in
figs.~\ref{fig1}(b)-(d). In the simulations we set $m=1$ and $T=1$
(the thermodynamic temperature), and use three different values of
$\alpha$: $\alpha=0.1$ [fig.~\ref{fig1}(b)], $\alpha=1$
[\ref{fig1}(c)], $\alpha=10$ [\ref{fig1}(d)]. Based on the simulation
results for this model in ref.~\cite{gjf3}, we restrict the
simulations to the range $0<dt\leq 1$ at which the GJF algorithm
exhibits accurate configurational sampling. For $dt>1$, discrepancies
between $T_c$ and $T$ in the low friction simulations become
noticeable (relative error $>6\%$). We measure the configurational
temperature, $T_c$ [denoted by black circles in
  figs.~\ref{fig1}(b)-(d)], and three kinetic temperatures $T_k^v$
(red squared), $T_k^u$ (green diamonds), $T_k^w$ (blue triangles)
corresponding, respectively, to the in-site velocity $v^n$ defined in
the original velocity-Velret GJF algorithm [Eq.~(\ref{eq:gjfv})], and
the two half-step velocities $u^{n+1/2}$ [Eq.~(\ref{eq:gjfv2})] and
$w^{n+1/2}$ [Eq.~(\ref{eq:gjfvlf})] introduced in the GJF-F and
GJF-2GJ leap-frog formulations of the GJF algorithm. The simulation
results in figs.~\ref{fig1}(b)-(d) clearly demonstrate the difference
between the in-site and half-step velocity variables. While the
kinetic temperature associated with the former tends to decrease with
$dt$, the kinetic energy of the latters remains extremely close to the
average thermodynamic kinetic energy ($0.99<T_k/T<1.015$) for any time
step within the range simulated herein. These results corroborate the
intuition from the harmonic oscillator analysis that the half-step
discrete-time velocity variables $u^{n+1/2}$ and $w^{n+1/2}$ are
robust to time step variations also in non-linear systems. This
conclusion agrees with the recent findings reported in
ref.~\cite{2gj}, where the robustness of the half-step velocity
$w^{n+1/2}$ was demonstrated in simulations of three-dimensional
Lennard-Jones systems.

\section{Summary}

We have introduced the GJF-F algorithm,
Eqs.~(\ref{eq:gjfv2})-(\ref{eq:gjfr2}), which is a new formulation of
the GJF algorithm for Langevin dynamics simulations. In this
formulation, the GJF thermostat is represented as a leap-frog scheme
with half-step velocity $u^{n+1/2}$.  In contrast to the in-site
velocity variable $v^n$ appearing in the original GJF algorithm, the
half-step velocity in the new GJF-F algorithm exhibits robustness to
time step variations when applied to the harmonic oscillator
problem. Computer simulations demonstrate that this feature of the
half-step velocity is also observed in strongly non-linear
systems. Thus, the newly-presented method allows for both accurate
configurational and kinetic sampling of canonical ensembles. This
makes the method very attractive for implementation in computer
simulations. On the one hand, it generates the same trajectories,
$\{r^n\}$, like the GJF algorithm and thus provides high quality  
configurational sampling with larger time steps compared to other
popular Langevin thermostats. On the other hand, it also provides
robust kinetic sampling, which offers a convenient way to assessing the
temperature of the simulated system via the average kinetic energy.

{\bf Acknowledgments:} I thank Niels Gr\o nbech-Jensen for stimulating
discussions, especially related to the differences between on-site and
half-step velocities. This work was supported by the Israel Science
Foundation (ISF) through Grant No. 991/17.


\begin{thebibliography}{99}

\bibitem{langevin:1908} P. Langevin, On the theory of Brownian motion,
  C. R. Acad. Sci. (Paris) {\bf 146}, 530 (1908).

\bibitem{risken:book} ] H. Risken, {\em The Fokker-Planck Equation}\/
  (Springer, Berlin, 1984).

\bibitem{pbs88} R. W. Pastor, B. R. Brooks, and A. Szabo, An analysis
  of the accuracy of Langevin and molecular dynamics algorithms,
  Mol. Phys. {\bf 65}, 1409 (1988).

\bibitem{bbk84} A. Br\"{u}nger, C. L. Brooks, and M. Karplus,
  Stochastic boundary conditions for molecular dynamics simulations of
  ST2 water, Chem. Phys. Lett.  {\bf 105}, 495 (1984).

\bibitem{ss78} T. Schneider and E. Stoll, Molecular-dynamics study of
  a three-dimensional one-component model for distortive phase
  transitions, Phys. Rev. B {\bf 17}, 1302 (1978).

\bibitem{vGB82} W. F. van Gunsteren and H. J. C. Berendsen, Algorithms
  for Brownian dynamics, Mol. Phys. {\bf 45}, 637 (1982).

\bibitem{wang03} W. Wang and R. D. Skeel, Analysis of a few numerical
  integration methods for the Langevin equation, Mol. Phys. {\bf 101},
  2149 (2003).

\bibitem{lm12} B. Leimkuhler and C. Matthews, Rational construction of
  stochastic numerical methods for molecular sampling,
  Appl. Math. Res. Express {\bf 2013}, 34 (2012).
  
\bibitem{gjf1} N. Gr\o nbech-Jensen, and O. Farago, A simple and
  effective Verlet-type algorithm for simulating Langevin dynamics,
  Mol. Phys. {\bf 111}, 983 (2013).

\bibitem{gjf2} N. Gr\o nbech-Jensen, N. R. Hayre, and O. Farago,
  Application of the G-JF discrete-time thermostat for fast and
  accurate molecular simulations, Comput. Phys. Commun. {\bf 185}, 524
  (2014).

\bibitem{gjf3} E. Arad, O. Farago, and N. Gr\o nbech-Jensen, The G-JF
  thermostat for accurate configurational sampling in soft-matter
  simulations, Isr. J. Chem. {\bf 56}, 629 (2016).

\bibitem{verlet67} L. Verlet, Computer "experiments" on classical
  fluids. I. Thermodynamical properties of Lennard-Jones molecules,
  Phys. Rev. {\bf 159}, 98 (1967).

\bibitem{2gj} L. F. Gr{\o}nbech-Jensen and N. Gr{\o}nbech-Jensen,
  Accurate configurational and kinetic statistics in discrete-time
  Langevin systems, Mol. Phys. (2019);
  https://doi.org/10.1080/00268976.2019.1570369.
    
\end{thebibliography}
\end{document}